\documentclass[preprint,showpacs,showkeys,amsmath,amssymb,aps,doublespace]{revtex4}
\usepackage[dvips]{graphicx}
\usepackage{setspace}
\usepackage{amsmath}
\usepackage{amsfonts}
\usepackage{amssymb,color}

\begin{document}

\title{Complete high-precision entropic sampling}

\author{Ronald Dickman\footnote{e-mail address: dickman@fisica.ufmg.br} and
A. G. Cunha-Netto\footnote{e-mail address: agcnetto@fisica.ufmg.br}}
\affiliation{Departamento de F\'{\i}sica, Instituto de Ci\^{e}ncias Exatas,
and National Institute of Science and Technology
for Complex Systems,
Universidade Federal de Minas Gerais, C.P. 702, 30123-970 Belo Horizonte, Minas Gerais, Brazil}

\begin{abstract}
Monte Carlo simulations using entropic sampling to estimate the number of
configurations of a given energy are a valuable alternative to traditional methods.
We introduce {\it tomographic} entropic sampling, a scheme which uses multiple studies,
starting from different regions of configuration space, to yield
precise estimates of the number of configurations
over the {\it full range} of energies, {\it without} dividing the latter into subsets or windows.
Applied to the Ising model on the square lattice,
the method yields the critical temperature to an accuracy of about 0.01\%, and critical
exponents to 1\% or better.  Predictions for systems sizes $L=10$ - 160, for the temperature of the
specific heat maximum, and of the specific heat at the critical temperature, are in very
close agreement with exact results.  For the Ising model on the simple cubic lattice the critical
temperature is given to within 0.003\% of the best available estimate;
the exponent ratios $\beta/\nu$ and $\gamma/\nu$ are given to within about 0.4\%
and 1\%, respectively, of the literature values.
In both two and three dimensions, results for the {\it antiferromagnetic} critical point are fully consistent
with those of the ferromagnetic transition.
Application to the lattice gas with nearest-neighbor exclusion on the square lattice again yields
the critical chemical potential and exponent ratios $\beta/\nu$ and $\gamma/\nu$ to good precision.
\end{abstract}


\maketitle

\section{Introduction}

A simple and reliable Monte Carlo method,
capable of estimating thermodynamic properties at any desired temperature using a single
simulation, has been sought for many years \cite{lee93,broad,tmmc,flat}.
Such a method could in principle offer a great advantage over
Metropolis Monte Carlo and cluster algorithms, which require a separate study for each
temperature of interest.  The information required to study the full temperature range
is contained in the number of configurations $\breve{\Omega}(E,L)$
as a function of energy $E$, and associated microcanonical averages [for example, the
mean magnetization $\langle m \rangle (E,L)$], on a lattice of $L^d$ sites.
Wang-Landau sampling (WLS) \cite{landau,landaupre} is currently the most
widely used such {\it entropic sampling} method.  Various improvements to the basic
WLS scheme have been suggested \cite{nfold,schulz,belardinelli,adaptive-window}.
In WLS of larger systems the range of energies (or of other relevant quantities)
is divided into subsets and each such ``window" is sampled separately.
The latter procedure has been found to distort the estimates for
$\breve{\Omega}(E,L)$ in some cases \cite{schulz,adaptive-window}.

Despite the broad interest in WLS, relatively few works have been published in
which critical exponents are determined using the method.  Important exceptions
to this general trend are the studies of Malakis and coworkers \cite{mal04,mal05,mal08,mal09,mal10},
who use a modified WLS procedure to find the critical exponents of several pure and disordered
spin models.  Their method involves a two-stage process, first using WLS with the energy range
divided into windows to identify the critical energy subspace
(i.e., the range of energies needed to sample the critical region adequately), followed
by WLS restricted to the critical range.  In Ref. \cite{nnelg} the present authors
obtained fair results for the critical exponents of the lattice gas with nearest-neighbor
exclusion, using WLS with adaptive windows.

Here we describe a high-precision entropic simulation method that
samples the full configuration space without using windows.
Two fundamental aspects of the method are: (1) successive refinement of an initial
estimate for the $\breve{\Omega}(E,L)$ on the basis of simulations starting from
a diverse set of initial configurations, and (2) use of the final result at size $L$ to
generate a good preliminary estimate for $\breve{\Omega}(E,L')$ for size $L' > L$.
After describing the method in Sec. II, we present, in Sec. III, applications to the Ising model on the
square and simple cubic lattices, and to the lattice gas with nearest-neighbor exclusion.
A summary and conclusions are given in Sec. IV.

\section{Method}

Consider a statistical model with a discrete configuration space, and let $\vartheta$ denote a variable
(or set of variables) characterizing each configuration, such as energy or particle number.  For a given
system size, knowledge of the number $\breve{\Omega}(\vartheta)$ of configurations,
for all allowed
values of $\vartheta$, permits one to evaluate the partition function and associated thermal averages
for any desired temperature.
Entropic sampling methods use Monte Carlo simulation to furnish estimates of the configuration
numbers, which we denote by $\Omega(\vartheta)$, reserving $\breve{\Omega}(\vartheta)$ to denote
the {\it exact} values, which are in general unknown.
The basic idea of entropic sampling is as follows.  Let $E({\cal C})$ denote the energy of
configuration ${\cal C}$.  Consider a simulation method that generates
configurations from the probability distribution

\begin{equation}
P({\cal C}) \propto \frac{1}{\breve{\Omega}[E({\cal C})]},
\label{probdist}
\end{equation}
i.e., the probability of configuration ${\cal C}$ is inversely proportional
to the number of configurations having the same energy.  This
could be done using an acceptance probability

\begin{equation}
p({\cal C'})= \min\left[ \frac{\breve\Omega(E)}{\breve\Omega(E')},1 \right].
\label{paccept}
\end{equation}

\noindent (As usual,
when the new configuration ${\cal C'}$ is rejected, the current configuration
${\cal C}$ is counted again in the sample.)
Let $H(E)$ be the number of times energy $E$ occurs in the sample.  In a
simulation that follows Eq. (\ref{probdist}),
$\langle H(E) \rangle = $ {\it Const.}, independent of $E$.
Of course, the $\breve\Omega (E)$ are in general unknown.  If we run the simulation
using a set of estimated values, $\Omega(E)$, in place of $\breve\Omega(E)$
in Eq. (\ref{probdist}), then the better the estimate,
the more uniform the histogram $H(E)$ should be.  The expected number of
visits to energy $E$ when we use an estimate $\Omega (E)$ to define $p_a({\cal C})$
is

\begin{equation}
\langle H(E) \rangle \propto \frac{\breve\Omega(E)}{\Omega(E)}
\label{HE}
\end{equation}

Suppose we begin using a {\it guess}, $\Omega_0(E)$, for $\breve\Omega(E)$.  Simulating
for a sufficiently long time, so that $H(E) \approx \langle H(E) \rangle$ for all
energies $E$, we may refine our guess, using Eq. (\ref{HE}) to write

\begin{equation}
\Omega_{1} (E) = \frac{H_1 (E)}{\overline{H}_1} \Omega_0 (E)
\label{refine}
\end{equation}

\noindent where $H_1 (E)$ is the number of times energy $E$ occurs in the simulation,
and $\overline{H}_1$ is the average of $H_1 (E)$ over all allowed energies.  Equation (\ref{refine})
is the basis of our simulation method.  The idea is to iterate this procedure, starting from an
initial guess, enough times so that the final approximation, $\Omega_N$, is very close to the
true distribution.  (During the final iteration, the microcanonical averages required to calculate
thermal averages are also determined.)

For the method to yield a reliable estimate for $\breve\Omega(E)$, it is essential that the
space of configurations be sampled long enough, and broadly enough, that at each step $j = 1,...,N$,
the observed histogram $H_j(E)$ is close to its expected value, that is,
$H_j (E) \propto \breve\Omega(E)/\Omega_{j-1} (E)$, for all allowed values of the energy.
If this condition is satisfied at each iteration $j$, we expect the estimates $\Omega_{j-1} (E)$
to approach the exact values.

The preceding discussion suggests that, to converge, the configuration space must be sampled
well at each iteration.
Our numerical experiments confirm that better results are obtained using a relatively small number
of iterations, with a relatively large number of lattice updates (LUDs)~\cite{lud} than with a
larger number of iterations, each having a
smaller number of updates, as shown in the following section.
In the studies reported below, for example, we use $N=5$ steps,
each consisting of a set of ten simulations, each of which runs for $N_U = 10^7$ LUDs.

To ensure
adequate sampling of the full range of energies, the simulations in a given iteration
use a variety of initial configurations.  For the Ising model on the square and
simple cubic lattices, for which the number of
configurations is invariant under (1) inversion of all spins, and (2) inversion of all spins on
{\it one sublattice}, we use the following set of initial conditions: (1 and 2) random spin orientations;
(3) all spins up; (4) all spins down; (5) all but two neighboring spins up;
(6) all but two neighboring spins down; (7 and 8) all
spins on one sublattice up, the rest down; (9 and 10) configurations (7) or (8) with
a pair of neighboring spins inverted.  Thus there are four low-energy initial configurations, four of
high energy (but of low energy for the antiferromagnetic model), and two that have $E \approx 0$.
Taking a certain liberty with terminology, we call this approach {\it tomographic sampling} of
the distribution $\Omega_j$, since studies using different starting points are pooled at each
iteration.

A key question concerns the initial estimate, $\Omega_0$, for the number of configurations.  Here
it is important to note that, in terms of the energy density $e = E/L^d$, the
microcanonical entropy density,

\begin{equation}
s(e,L) \equiv \frac{1}{L^d} \ln \breve\Omega (L^d e,L)
\label{defs}
\end{equation}

\noindent depends on system size $L$ rather weakly.  Thus the final result for $s(e,L)$ serves as
a rather good first approximation for a larger system size, $L'$.
We use a simple initial guess (for example, $\Omega (E)$ uniform over the set of allowed energies,
or a mean-field approximation) for some small initial system size.  For small
$L$ the simulation rapidly converges to a good estimate of $\breve\Omega (E,L)$, and the resulting
entropy density $s(e,L)$ is then used to generate the initial guess $\Omega_0 (E,L')$ for the
next system size in the sequence.

\subsection{Implementation}

All the studies reported here use periodic boundaries; other boundary conditions
are readily implemented.
The Ising spin configurations are classed by
the {\it bond number} $n$, i.e., the number of nearest-neighbor
pairs of spins having the same orientation.  On the square
lattice the
allowed values are $n = 0, 4, 6,..., n_{max}-4, n_{max}$ with $n_{max} = 2 L^2$.
(The energy is $E= 2(L^2 - n)$.)
To begin, we set the initial distribution $\Omega_0 (n)$.
Due to the extremely large values these numbers may take (already $\sim {\cal O}(10^{30})$ for
$L=10$), they are represented in the form of their natural logarithm.
Subsequently, the simulation procedure is called $N$ times.
(In most of the studies reported here, $N=5$.)  The simulation uses a single-spin-flip dynamics.
The acceptance probabilities $\Omega_j (n)/\Omega_j (n')$ (for a transition from energy $n$ to $n'$),
are stored in a look-up table.  (The allowed values of
$\Delta n$ are 0, $\pm 2$, and $\pm 4$.)  In addition to storing the orientation of
each spin, we store the number of neighbors having the same spin orientation, as this
facilitates evaluation of $\Delta n$.

Each time the simulation subroutine is called, the histogram $H_j (n)$ is set to zero.
The latter then accumulates the number of visits to each class $n$ over $N_{sim} = 10$
simulations (with varied initial configurations), each consisting of
$N_U$ LUDs.  (In principle, these studies could be carried out in
parallel, on separate processors.) Once all $N_{sim}$ studies are done we update $\Omega (n)$
as per Eq.~(\ref{refine}).  The acceptance probabilities are recalculated using the new estimate, and
we proceed to the next iteration, for a total
of $N$ iterations.  Each time a configuration ${\cal C}$ is generated, be it a new one (following
an accepted spin flip) or a repetition (if the new configuration is rejected), we update
the running sums used to calculate the microcanonical averages of the absolute magnetization
$|M|$, as well as $M^2$ and $M^4$.  The resulting averages are saved, along with
$\ln \Omega_j (n)$, at the end of each iteration.  (Note however
that only the results from the final iteration, $j=N$, are used to calculate the thermodynamic properties
reported below.)

As mentioned above, the initial distribution $\Omega_0 (n,L)$ is generated on the basis of
the final estimate for a smaller system size.  For the smallest size ($L=10$ in the case of
the square lattice) we use a simple approximation for the entropy density,

\begin{equation}
s(\rho) \approx - \ln 2 (\rho - 1)^2
\label{srhoinit}
\end{equation}

\noindent where $\rho = n/L^2$.  This is motivated by the fact that $\breve\Omega (n=L^2) \sim 2^{L^d}$
whereas $\breve\Omega (n=2L^d) = \breve\Omega (n=0) = 2$, and that $s(\rho) $ is symmetric
about $\rho = 1$.  (For convenience we subtract $\ln 2$ from the
original definition, $s = [\ln \breve\Omega (n)]/L^2$.)  For $L=10$, the distribution after the
first iteration,
$\Omega_1 (n)$, is already very close to the final one, showing rapid convergence.

The quantity
$h(\rho) \equiv [H(\rho)-\overline{H}]/\overline{H}$ characterizes the
relative deviation of the histogram from a uniform distribution; its evolution at successive iterations
is illustrated in Fig. \ref{hrho}.  The result of the first iteration (starting from the parabolic
initial approximation) shows relatively large variations, while the subsequent histograms are all
flat on the scale of this figure. The detailed evolution
in the subsequent iterations is shown in the inset; the final histogram is flat to about 99.5\%.

\begin{figure}[!htb]
\caption{(Color online) Ising model, square lattice: relative deviation of histogram,
$h(\rho) \equiv [H(\rho)-\overline{H}]/\overline{H}$, versus bond density
$\rho$ for $L=10$.  The black curve is the result of the first
iteration, while the horizontal line corresponds to the four subsequent iterations.
Inset: Detail of iterations 2 - 5.
}
\label{hrho}
\end{figure}

For the passage from one system size ($L$), to the next ($L'$), we require an algebraic approximation for
the entropy density $s(\rho,L)$.  Given that $\breve\Omega (n) = \breve\Omega (n_{max} -n)$,
we first symmetrize the result about $\rho=1$, and average over pairs of neighboring values
to suppress small systematic oscillations in $s$.  (The latter are visible in Fig.~2, near $\rho=0$ and
$\rho = 2$.)
This defines the symmetrized and (slightly) smoothed entropy density,
\begin{equation}
\tilde{s} (\rho') \equiv \frac{1}{4} \left[ s(n/L^2) + s((n-1)/L^2) + s(2-n/L^2) + s(2-(n-1)/L^2) \right]
\end{equation}
where $\rho' = (2n-1)/(2L^2)$.
For smaller sizes, $\tilde{s}$ is well approximated by a polynomial fit in even powers of
$x \equiv \rho - 1$. For larger systems, however, maintaining a good fit requires increasing
the order of the polynomial.  The fit tends to be worst near $x = \pm 1$.  (In fact, one
may anticipate a weak singularity, $s \sim |1-x| \ln |1-x|$ at these limits.)  We find that,
given the symmetry $s(-x) = s(x)$, a more
convenient alternative is a Fourier cosine series:

\begin{equation}
s(x) - s (1) = \sum_{j=0}^J a_j \cos \left[\frac{(2j+1) \pi x}{2} \right] \equiv f(x)
\label{sfourier}
\end{equation}

\noindent The expansion coefficients are readily calculated using the discrete approximation
to the standard integral expression; the quality of the fit varies considerably
with the number of terms in the series.  We calculate the
maximum absolute deviation $\delta \equiv \max_x |s(x) - f(x)|$ as a function of
the number of terms, and use the value that minimizes $\delta$.
For example, for $L=120$ (square lattice), we obtain
the smallest $\delta$ ($2.7 \times 10^{-5}$) for a series with $J=9151$.
We use the Fourier series for
all studies of the two- and three-dimensional Ising models reported below.

The execution time (for a given number of iterations, initial configurations, and lattice updates)
scales as the number of lattice sites.
The largest studies reported here (for system size $L=160$)
required approximately one week on a processor with a speed of 2.8 GHz.

\section{Results}

\subsection{Ising model on the square lattice}

We apply tomographic sampling to the Ising model on the square lattice, for system
sizes $L=10$, 20, 40, 80, 120, and 160.  The procedure is as outlined above, except that for
the two largest sizes, following the five studies using $N_U = 10^7$ lattice updates, a sixth study using
$N_U = 2 \times 10^7$ is performed.  In all cases, the results are calculated on the basis
of the final study.  Means and uncertainties are calculated using five or in some cases
six independent simulations.

The number of configurations $\Omega (\rho)$ takes its maximum at $\rho=1$ and is (except for
fluctuations) symmetric about this point.  In Fig. \ref{srho} we plot the microcanonical entropy density

\begin{equation}
s(\rho) = \frac{1}{L^2} \ln \left( \frac{\Omega(\rho)}{\Omega(1)} \right)
\label{eqsrho}
\end{equation}

\noindent for several system sizes.  At this scale, finite-size effects are only evident near the limiting
bond densities.

\begin{figure}[!htb]
\caption{(Color online) Microcanonical entropy density $s(\rho)$ versus
$\rho$ for system sizes $L=10$, 20, 40, 80, and 160 (upper to lower).  The inset is a detailed view
near $\rho=2$.
}
\label{srho}
\end{figure}

To perform a quantitative test of the method we compare results for the specific heat,
magnetization, susceptibility, and reduced fourth cumulant with known values and with
finite-size scaling theory \cite{fisher,barber}.  We
define the specific heat per site as $c = \mbox{var}(E)/(L^d k_B T^2)$, where $E$ is the
total energy.  (From here on we employ units such that Boltzmann's constant $k_B = 1$.) Figure
\ref{c80c} shows $c(T)$ (for $L=80$) as obtained in five independent simulations.
The curves are indistinguishable on the scale of the main graph, but a detail of the
critical region does reveal some scatter in the specific heat maxima and the temperatures
at which they occur.  From the analysis of Ferdinand and Fisher \cite{ferdinand} one knows
that on a square lattice of $L \times L$ sites (with periodic boundaries), the specific heat
takes its maximum value at temperature $T(c_{max}) \simeq T_c(1 + 0.3603/L)$, where
$T_c \simeq 2.269185$ is the critical temperature.  Combining the results of Refs.
\cite{ferdinand} and \cite{salas} one also has an expansion for the specific heat
at the critical temperature:

\begin{equation}
c(T_c,L) = A_0 \ln L + C_0 \;+\; \frac{C_1}{L} \;+\; \frac{C_2}{L^2} \;+\; \frac{C_3}{L^3} \;+\; \cdots
\label{cexp}
\end{equation}
\vspace{.5em}

\noindent where $A_0 = (2/\pi) [\ln(1 + \sqrt{2})]^2 \simeq 0.494358$ and, for an $L \times L$ lattice
with periodic boundaries, $C_0 \simeq 0.138149$, $C_1 \simeq -0.170951$, $C_2 \simeq 0.018861$,
and $C_3 \simeq 0.056765$.  In Fig. \ref{ccrit} we plot $c(T_c)$ versus $\ln L$; a least-squares
linear fit yields $c(T_c) = 0.498(2) \ln L$, i.e., about 1\% above the exact amplitude.
A detailed comparison with theoretical predictions
is given in Table \ref{tab1}; simulation results agree with theory to within uncertainty.
Linear extrapolation of the simulation results for $T(c_{max})$ (for $L \geq 20$) versus $1/L$
yields $T_c = 2.26966(8)$, that is, about 0.02\% above the exact value of 2.269185... \cite{kramers}.
A similar analysis of
the energy per site at $T_c$ yields $e_c = -1.4147(5)$, consistent with the exact
value, $e_c = -\sqrt{2}$ \cite{kramers}.
As noted above, our sampling method represents the antiferromagnetic (AF) critical point with
the same precision as the ferromagnetic one.  We have verified that
$T(c_{max})$, $c_{max}$, and $c(T_c)$
associated with the AF transition agree to within
uncertainty with the values quoted in Table \ref{tab1} for the ferromagnetic transition.

\begin{figure}[!htb]
\caption{(Color online) Specific heat per site $c$ versus
temperature for $L=80$ in five independent simulations.
}
\label{c80c}
\end{figure}

\begin{figure}[!htb]
\caption{(Color online) Specific heat per site at the critical temperature $c_c$ versus
$\ln L$.  The slope of the regression line is 0.498(2).
}
\label{ccrit}
\end{figure}

\begin{table}[h]
\begin{center}
\begin{tabular}{|r|c|c|c|c|} \hline
$L$  & $T(c_{max})$ (th) & $T(c_{max})$ (sim) & $c(T_c)$ (th) & $c(T_c)$ (sim) \\ \hline\hline
 10  & 2.35              & 2.34450(6)       & 1.2600        & 1.2597(3)     \\
 20  & 2.310             & 2.30806(8)       & 1.6112        & 1.6121(5)     \\
 40  & 2.290             & 2.2889(2)        & 1.9582        & 1.9573(18)    \\
 80  & 2.2794            & 2.2793(2)        & 2.3031        & 2.302(3)      \\
120  & 2.2760            & 2.2761(1)        & 2.5043        & 2.504(5)      \\
160  & 2.2743            & 2.2743(2)        & 2.64695       & 2.655(7)      \\
\hline
\end{tabular}
\end{center}
\caption{\sf Square lattice: comparison of theoretical and simulation results for the temperature
of the specific heat maximum, and for the specific heat at $T_c$.  Note that the theoretical
expressions for $T(c_{max})$ and $c(T_c)$ are subject to corrections of order $1/L^2$ and $1/L^4$, respectively.}
\label{tab1}
\end{table}

We turn now to the magnetization, the susceptibility, and Binder's cumulant.
The magnetization per site is given by

\begin{equation}
m(T,L) = \frac{1}{L^d Z(T,L)} \sum_E \Omega(E,L) e^{-\beta E} \langle |N_+ - N_-| \rangle(E,L)
\label{mag}
\end{equation}

\noindent where $\beta=1/k_B T$, $Z$ is the partition function, and $\langle |N_+ - N_-| \rangle(E,L) $
denotes the microcanonical average
of the absolute value of the total magnetization ($N_+$ is the number of sites with spin $\sigma = 1$).
Results for the magnetization are plotted in Fig. \ref{m2d}.  The uncertainties, which are not
visible on the scale of the main graph, are shown in the inset; as is to be expected, they
are most severe in the vicinity of the critical point.  We note however that in all cases,
the relative uncertainty in $m$ is less than 0.4\%.
The finite size scaling (FSS) relation $m(T_c,L) \sim L^{-\beta/\nu}$ permits
one to estimate the associated exponent ratio.  Using the data for all system sizes we
find $\beta/\nu = 0.1237(5)$, that is, 1\% smaller than the exact value of 1/8.

\begin{figure}[!htb]
\caption{(Color online) Magnetization per site versus temperature for system sizes
$L=10$, 20, 40, 80, and 160.
Uncertainties are plotted in the inset.
}
\label{m2d}
\end{figure}

The susceptibility $\chi = \mbox{var}(M)/(L^d \, T)$,
and its uncertainty are plotted in Fig. \ref{chi2d}.  The relative uncertainty in $\chi$ is largest
(about $1$\%) in the critical region, for $L=160$; for smaller systems and away from the critical
region, it is considerably smaller.
We list simulation results for the temperature of the susceptibility maximum, $T(\chi_{max})$,
and for the
susceptibility at $T_c$, in Table \ref{tab2}.  Extrapolating the values of $T(\chi_{max})$ we
find $T_c = 2.26926(10)$.
Analysis of the susceptibility data using the FSS relation
$\chi (T_c,L) \sim L^{\gamma/\nu}$ yields $\gamma/\nu = 1.754(2)$,
which is 0.2\% higher than the exact value of $7/4$.
It is interesting to note that, restricting the analysis to system sizes $20 \leq L \leq 120$,
we obtain the slightly superior estimates $\beta/\nu = 0.1240(8)$ and $\gamma/\nu = 1.748(2)$.

\begin{table}[h]
\begin{center}
\begin{tabular}{|r|c|c|} \hline
$L$  & $T(\chi_{max})$  & $\chi(T_c)$  \\ \hline\hline
 10  & 2.4770(4)        & 1.7894(4)            \\
 20  & 2.3720(1)        & 6.093(5)             \\
 40  & 2.3208(2)        & 20.44(8)             \\
 80  & 2.2949(2)        & 68.95(25)            \\
120  & 2.2864(1)        & 139.5(8)             \\
160  & 2.28196(14)      & 233.2(1.8)           \\
\hline
\end{tabular}
\end{center}
\caption{\sf Square lattice: simulation results for the temperature
of the susceptibility maximum, and for the susceptibility at $T_c$.
}
\label{tab2}
\end{table}

\begin{figure}[!htb]
\caption{(Color online) Susceptibility per site versus temperature for system sizes $L=10$, 20, 40, 80, and 160.
Uncertainties are plotted in the inset.
}
\label{chi2d}
\end{figure}

An independent estimate of the critical temperature is afforded by
Binder's reduced fourth cumulant \cite{binder_cum}, $Q_4 = 1 - \langle M^4 \rangle/(3 \langle M^2 \rangle^2)$.
Analysis the crossings of Binder's cumulant for sucessive pairs of
system sizes yields the temperatures and cumulant values
listed in Table \ref{tab3}.  Linear extrapolation of the results for the
last four pairs yields $T_c = 2.2694(2)$ and $Q = 0.610(1)$.  The latter
is in good accord with the literature value $Q=0.61071(2)$ \cite{kamieniarz}.  Pooling the
estimates for the critical temperature derived from the analysis
of the specific heat, the susceptibility, and the reduced cumulant, we
find $T_c = 2.2695(1)$, about 0.01\% above the exact value.

Our estimate for $T_c$ is, as noted, based on extrapolations
of three sets of size-dependent {\it pseudocritical} temperatures, i.e.,
$T(c_{max})$, $T(\chi_{max})$, and $T_\times$.  These quantities are plotted
versus $1/L$ in Fig.~\ref{tpscr}.  We analyzed the behavior of
the pseudocritical temperature associated with the susceptibility maximum,
which should follow $T(\chi_{max}) - T_c \sim L^{-1/\nu}$.
Using our estimate of $T_c = 2.2695$, a linear fit of
$\ln [T(\chi_{max}) - T_c]$ versus $\ln L$ yields $\nu = 0.99(1)$
consistent with the critical exponent
$\nu=1$ for the two-dimensional Ising model.

\begin{figure}[!htb]
\caption{(Color online) Two-dimensional Ising model:
pseudocritical temperatures (upper to lower) $T(\chi_{max})$,
$T(c_{max})$, and $T_\times$ versus inverse system size.  Error bars
are smaller than the symbols.
}
\label{tpscr}
\end{figure}

\begin{table}[h]
\begin{center}
\begin{tabular}{|r|c|c|} \hline
$L$, $L'$  & $T_\times$  & $Q_\times$  \\ \hline\hline
 10, 20    & 2.2635(1)   & 0.61372(10)            \\
 20, 40    & 2.2692(2)   & 0.61115(20)            \\
 40, 80    & 2.2689(4)   & 0.6115(10)             \\
 80, 120   & 2.2696(4)   & 0.6096(17)             \\
120, 160   & 2.2693(5)   & 0.611(2)               \\
\hline
\end{tabular}
\end{center}
\caption{\sf Ising model, square lattice: simulation results for the temperature
and the cumulant value at crossings of Binder's cumulant for pairs of
successive system sizes.
}
\label{tab3}
\end{table}

As noted above, we find it preferable to perform a relatively small number of long
iterations than a large number of shorter ones.  To illustrate this point, we performed
a series of studies using a total of $5 \times 10^8$ LUDS, divided into 5, 10, 25 and 50
iterations, for system size $L=20$.  In Fig.~\ref{urel} we plot the relative uncertainties
(estimated using the standard deviation calculated over a set of five independent
studies),
of $T(c_{max})$, $T(\chi_{max})$, and of the magnetization, susceptibility, and
Binder's cumulant, evaluated at the critical temperature.  As is clear from the figure,
the uncertainties decrease systematically as we reduce the number of iterations; the relative
uncertainty is roughly 5 -10 times larger using 50 iterations as compared to only five.
The inset, for $m_c$, illustrates the general trend: despite the monotonic variation of
the uncertainties, the estimates obtained using different values of
$N$ are mutually consistent.

\begin{figure}[!htb]
\caption{(Color online) Relative uncertainties in $T(c_{max})$ (diamonds), $T(\chi_{max})$ ($\times$),
$m_c$ (filled squares), $\chi_c$ (open squares), and $Q_c$ (circles) for $L=20$, versus number
of iterations $N$, in studies using a total of $N \times N_U = 5 \times 10^7$ lattice updates.
Inset: estimates for $m_c$ versus $N$.
}
\label{urel}
\end{figure}

\subsection{Ising model on the simple cubic lattice}

The procedure is essentially the same as for the square lattice.  We begin sampling at
$L=4$, using the simple initial guess for $s(\rho)$ given in Eq. (\ref{srhoinit}).
Subsequently we study system sizes $L=8$, 12, 16, 20, 24, 28, 32, and 36.
The smaller increment in $L$ (as compared with the two-dimensional case), is motivated
by faster convergence, and by the need to study a reasonably large number of system sizes
in the FSS analysis, which in this case includes a correction to scaling term.

\begin{figure}[!htb]
\vspace{-2em}

\caption{(Color online) Ising model, simple cubic lattice: specific heat per
site versus temperature for system sizes $L=8$, 12, 16, 20,
24, 28, 32, and 36.
}
\label{spht3d}
\end{figure}

\begin{figure}[!htb]
\vspace{-2em}

\caption{(Color online) Ising model, simple cubic lattice: pseudocritical
temperatures $T(c_{max})$ (lower) and $T(\chi_{max})$ (upper) versus $1/L^{1/\nu}$.
Inset: temperatures $T_\times$ of Binder cumulant crossings versus
$1/\overline{L}$ (here $\overline{L}$ denotes the geometric mean of the system sizes).
}
\label{tpssc}
\end{figure}

The specific heat in the critical region is shown in Fig. \ref{spht3d}; the relative
uncertainty in these results is at most 0.5\%.
We estimate the critical temperature on the basis of $T(c_{max})$ and $T(\chi_{max})$ (see Fig.~\ref{tpssc}).
Using the values of $T(c_{max})$ for $L=12$ - 36, a quadratic least-squares fit versus $1/L^{1/\nu}$
yields $T_c = 4.5116(1)$; a similar analysis using $T(\chi_{max})$ yields 4.5114(1).
(We use the literature value, $\nu = 0.6301(8)$ \cite{blote95}.)
These results are in good accord with the best
estimate of $T_c = 4.511528(6)$ \cite{talapov}.
The energy per site at the critical point, $e_c(L)$, is essentially linear when plotted
versus $1/L^{1/\nu}$; extrapolation to infinite size yields $e_c = -0.9928(2)$, which
compares well with the series-expansion estimate, $e_c = -0.9922$ \cite{sykes}.
For each system size,
the values of $c_{max}$ and
$T(c_{max})$ associated with the antiferromagnetic critical point are fully consistent
with the corresponding ferromagnetic values.

\begin{figure}[!htb]
\vspace{-2em}

\caption{(Color online) Ising model, simple cubic lattice: magnetization (lower),
specific heat (middle), and susceptibility (upper) evaluated at the critical temperature
versus system size.  Error bars are smaller than symbols.
}
\label{fss3d}
\end{figure}

The critical specific heat, susceptibility, and magnetization are plotted
versus system size (on log scales)
in Fig.~\ref{fss3d}.  In light of the slight curvature evident in these plots, and
given the smaller system sizes used in the three-dimensional studies, we found it useful to
include a correction to scaling in the FSS analysis.  For example, we fit the data for the
magnetization at $T_c$ using

\begin{equation}
\ln m(T_c,L) = -\frac{\beta}{\nu} \ln L + c L^{-y_i} + \mbox{Const.} \equiv f_L + \mbox{Const.}
\label{betafit}
\end{equation}

\noindent where $y_i$, the dominant irrelevant scaling exponent, is taken as 0.8.
(Note that the constant term is not a fitting parameter, as we determine the values of
$\beta/\nu$ and $c$ by minimizing the variance between the simulation data and $f_L$.)
The best-fit parameters
for system sizes $L=8$ - 36
are $c = -0.064$ and $\beta/\nu = 0.521(12)$.
The uncertainty $\sigma_{\overline{\beta}}$ in $\overline{\beta} \equiv \beta/\nu$
is calculated as follows.  Let $Y_L \equiv \ln m(T_c,L) - f_L$, and let $\sigma_Y^2$ denote
the variance of the $Y_L$, for parameters $\overline{\beta}$ and $c$ chosen to minimize this
variance.  Further, let $\sigma_L^2$ denote the statistical uncertainty in the simulation
result for $\ln m(T_c,L)$, and define $\overline{\sigma}_L^2 \equiv \sigma_L^2 + \sigma_Y^2$.
Then we take

\begin{equation}
\sigma_{\overline{\beta}}^2 = \sum_L \overline{\sigma}_L^2
\left(\frac{\partial \overline{\beta}}{\partial \ln m(T_c,L)} \right)^2,
\label{uncbeta}
\end{equation}

\noindent where the derivatives are determined numerically.  The uncertainties in the other exponent
ratios are estimated in a similar manner.

For the susceptibility, we expect the dominant correction
to FSS to be $\propto L^{-y_2}$ with $y_2 \simeq 1.96$ \cite{blote95}, and write

\begin{equation}
\ln \chi(T_c,L) = \frac{\gamma}{\nu} \ln L + c' L^{-y_2} + \mbox{Const.}
\end{equation}

\noindent In this case we find $\gamma/\nu = 1.987(4)$.  (Using a correction to
scaling term $\propto 1/L$ we instead obtain
$\gamma/\nu = 1.955(11)$.)  A similar analysis of the specific heat data
(for system sizes $L=12$ - 36, using a correction term $\propto L^{-\alpha/\nu}$)
yields $\alpha/\nu = 0.161(3)$.  Compared with the literature values \cite{talapov},
$\beta/\nu = 0.519(2)$, $\gamma/\nu = 1.963(3)$, and $\alpha/\nu = 0.174(4)$, our results for
exhibit errors of 0.4\%, 1.2\%, and 7.5\%, respectively.  (We note, however, that determination of $\alpha/\nu$
via simulation is in general a difficult task.)

Extrapolation of the crossings of Binder's cumulant (see Fig.~\ref{tpssc}, inset),
yields $T_c = 4.5124(16)$, which, while
less precise than the estimates cited above, is consistent with the best estimate for
the critical temperature.  The asymptotic value of Binder's cumulant obtained from the crossings is
$Q_4 = 0.47(1)$, while extrapolation of the cumulant values $Q_4 (T_c,L)$ yields 0.467(1).  The
reference value for the three-dimensional Ising model is $Q_4=0.465(3)$ \cite{blote95}.

Entropic sampling is particularly advantageous for data collapse analyses,
as it furnishes thermodynamic quantities as continuous functions of temperature.
FSS predicts that magnetization curves for diverse system sizes should fall on
universal curves (for $T>T_c$ and $T<T_c$) when plotted in the form
$L^{\beta/\nu} m (L^{1/\nu}|T-T_c|)$. Figure~\ref{cm3d}
shows a near-perfect collapse of the magnetization data for eight system sizes ($L=8$-36); the
associated exponents are $\nu=0.62$ and $\beta/\nu = 0.512$.  Figure~\ref{cchi3d} is a similar
plot for the susceptibility, again using $\nu=0.62$, and $\gamma/\nu = 1.99$.  In this case the
data for $T>T_c$ collapse perfectly while those for $T<T_c$ approach a common scaling function
with increasing system size.  (The downward curvature in these plots signals the
low-temperature boundary of the scaling region.)

\begin{figure}[!htb]
\vspace{-2em}

\caption{(Color online) Ising model, simple cubic lattice: scaled magnetization
$m^* \equiv L^{\beta/\nu} m$ versus scaled temperature $t^* = L^{1/\nu} |T-T_c|$
for eight system sizes ($L=8$-36), using $\nu=0.62$ and $\beta/\nu = 0.512$.
}
\label{cm3d}
\end{figure}

\begin{figure}[!htb]
\vspace{-2em}

\caption{(Color online) Ising model, simple cubic lattice: scaled susceptibility
$\chi^* \equiv L^{-\gamma/\nu} \chi$ versus scaled temperature $t^* = L^{1/\nu} |T-T_c|$
for eight system sizes ($L=8$-36), using $\nu=0.62$ and $\gamma/\nu = 1.99$.
}
\label{cchi3d}
\end{figure}

\subsection{Lattice gas with nearest-neighbor exclusion}

We apply our entropic sampling to the lattice
gas with nearest neighbor exclusion (NNE) on the square lattice \cite{runnels,blote02,nnelg},
using entropic sampling
to estimate the number of distinct configurations
$\breve{\Omega}({\cal N},L)$ with ${\cal N}$ particles satisfying the NNE condition,
on an $L \times L$ lattice with periodic boundaries. The grand partition function is

\begin{equation}
\Xi(z,L)=\sum_{{\cal N}=0}^{{\cal N}_{max}}z^N\breve{\Omega}({\cal N},L),
\end{equation}
where $z=e^{\mu}$ is the fugacity, $\mu\equiv\widehat{\mu}/k_BT$
($\widehat{\mu}$ denotes the chemical potential), and ${\cal N}_{max}$ is the maximum
possible number of particles, equal to $L^2/2$ on the square lattice.
(In what follows we refer to $\mu$ as the chemical potential.)
The model exhibits an Ising-like phase transition on bipartite lattices;
in the ordered phase a majority
of the particles occupy one of the sublattices.
The associated order parameter is
the difference between the occupancies of sublattices A and B:
\begin{equation}
\phi=\frac{1}{{\cal N}_{max}}\left\langle \left\vert \sum_{\bf x \in {\rm A}}\sigma_{\bf x}-
\sum_{\bf x \in {\rm B}}\sigma_{\bf x} \right\vert \right\rangle,
\label{eq:ord_par}
\end{equation}
where $\sigma_{\bf x}$ is the indicator variable for occupation of site {\bf x}.

We begin our study with a small system $(L=8)$, using a flat
initial distribution, $\ln \Omega_0 =0$.
Following the procedure described in Sec. II,
we perform $N=5$ iterations, with increasing numbers
of lattice updates \cite{note_luds}.
We use ten initial configurations at each iteration: five corresponding to an empty lattice, and
five in which one sublattice is fully occupied.
Here each trial move is either an insertion or a removal of a particle; target sites
are chosen at random. For system sizes $L' > 8$ we construct the initial guess
for $\breve{\Omega}({\cal N},L')$ using
a tenth-degree polynomial fit
to $s(\rho,L) = [\ln \breve\Omega ({\cal N},L)]/L^2$,
where $L$ is the previous size studied and $\rho = {\cal N}/L^2$.
We study six sizes in the range $8\leq L \leq 120$.

Figure~\ref{fig:fi_ki} shows the susceptibility,
$\chi(\mu)=L^2(\left\langle \phi^2\right\rangle_{\mu}
-\left\langle \phi\right\rangle_{\mu}^2)$, as a function
of the chemical potential for different system sizes; the inset is a similar plot of
the order parameter.
Estimates for the critical chemical potential $\mu_c$ are obtained
via analysis of the maxima of the susceptibility and of the compressibility,
$\kappa(\mu)=L^2(\left\langle \rho^2\right\rangle_{\mu}-
\left\langle \rho\right\rangle_{\mu}^2)/\left\langle \rho\right\rangle_{\mu}^2$.
Extrapolation versus $1/L$ yields
$\mu_{c,\chi}=1.3359(3)$ and $\mu_{c,\kappa}=1.344(2)$.
The former, more precise value, is about 0.14\% higher than the literature value.
FSS analysis of the susceptibility maxima
yields $\gamma/\nu=1.750(2)$, consistent with the exact value
for the two-dimensional Ising model.

Using the high-precision result for the critical chemical potential
obtained by Guo and Bl\"ote \cite{blote02}, $\mu_c=1.33401510027774(1)$,
we calculate $\rho_c(L)$, $\phi_c(L)$
and $Q_{c}(L)$, where $Q= \left\langle \phi^2\right\rangle^2/\left\langle \phi^4 \right\rangle$ is
related to Binder's reduced cumulant \cite{binder_cum}. Linear
extrapolation of these quantities versus $1/L$ yields $\rho_{c}=0.36773(1)$, $\beta/\nu=0.1247(3)$
and $Q_c=0.8565(5)$. These results are in very good agreement with the
literature values \cite{kamieniarz,blote02}.
(Using our own less accurate estimate, $\mu_{c}=1.3359(3)$ we obtain $\beta/\nu = 0.122(2)$ and
$\rho_{c}=0.36815(3)$.)
We summarize our main results for the lattice gas and compare them with reference values in Table IV.

\begin{figure}[!htb]
\caption{(Color online) NNE lattice gas: susceptibility versus chemical potential.
Inset: order parameter versus chemical potential,
system sizes as
indicated. Error bars are smaller than the symbols.}
\label{fig:fi_ki}
\end{figure}

\vspace{1cm}

\begin{table}[thb!]
\singlespacing
\centering
\begin{tabular}{lllll}
\hline
      ~~~~~ & ~~ ES ~~ & ~~ AWWLS \cite{nnelg} ~~ & ~~  Literature values  \\ \hline
$\mu_{c,\chi}$ ~~ & ~~ $1.3359(3)$  ~~ & ~~ $1.330(1)$ ~~ & ~~ $1.33401510027774(1)$\footnotemark[1] \\
$\mu_{c,\kappa}$ ~~ & ~~ $1.344(2)$ ~~ & ~~ $1.337(2)$ ~~ & ~~  \\
$Q_c$       ~~ & ~~ $0.8565(5)$\footnotemark[5] ~~ & ~~ $0.852(6)$   ~~ & ~~ $0.856$\footnotemark[2]; $0.855(1)$\footnotemark[3]; $0.85625(5)$\footnotemark[4]  \\
$\rho_{c}$ ~~ & ~~ $0.36773(1)$\footnotemark[5] ~~ & ~~ $0.36800(5)$\footnotemark[5] ~~ & ~~ $0.3677429990410(3)$\footnotemark[1] \\
$\gamma/\nu$ ~~ & ~~ $1.750(2)$ ~~ & ~~ $1.762(8)$    ~~ & ~~ $7/4$ (exact) \\
$\beta/\nu$ ~~ & ~~ $0.1247(3)$\footnotemark[5]  ~~ & ~~ $0.123(2)$\footnotemark[5]    ~~ & ~~ $1/8$ (exact) \\
\hline
\end{tabular}
\footnotetext[1]{Guo and Bl\"ote\cite{blote02}}
\footnotetext[2]{Burkhardt and Derrida\cite{burkhardt}}
\footnotetext[3]{Nicolaides and Bruce\cite{nicolaides}}
\footnotetext[4]{Kamieniarz and Bl\"ote\cite{kamieniarz}}
\footnotetext[5]{Values obtained using $\mu_{c}$ from \cite{blote02}.}
\footnotetext[6]{Using our best value for $\mu_{c}$ we obtain $\beta/\nu=0.122(2)$,
$\rho_{c}=0.36815(3)$ while the AW study yields $\beta/\nu=0.130(9)$ and $\rho_{c}=0.3675(5)$.}
\caption{\sf Critical values for the NNE lattice gas
obtained via Entropic Sampling (ES) and WLS with adaptive windows (AW).
The results of \cite{nicolaides} were obtained via Monte Carlo simulation,
those of \cite{blote02,burkhardt,kamieniarz}
via transfer-matrix analysis.}
\end{table}

\subsection{Critical singularity of the entropy density}

The critical point corresponds to a value of $\rho$ such that, in the thermodynamic limit,
the second derivative $s'' \equiv d^2 s/d \rho^2 = 0$.  Since direct estimation of the
second derivative of numerical data via finite differences is not a viable procedure, we
instead use a Gaussian filter,

\begin{equation}
s'' (\rho; \sigma) = \int_0^2 d \rho' s(\rho') g'' (\rho-\rho';\sigma)
\label{spp}
\end{equation}

\noindent where $g(\rho;\sigma)$ is a normalized Gaussian distribution
with mean zero and standard deviation $\sigma$.
(The filter width $\sigma$ is chosen large enough to suppress fluctuations
but small enough to maintain resolution; it ranges from about 5$\Delta \rho$ for $L=20$ to
about 50$\Delta \rho$ for $L=160$, where $\Delta \rho = 1/(2L^2)$ is the bond density increment.)
Figure \ref{spp2d}, for the square lattice, shows that $|s''|$
exhibits a sharp minimum, which appears to approach zero as $L$ increases.
The minimum near $\rho = 1.7$ corresponds to the ferromagnetic transition; note that
the expected (infinite-size) value is $rho_c = 1 - e_c/2 = 1+1/\sqrt{2} = 1.707\,107...$.
The minimum near
$\rho = 0.3$ corresponds to the {\it antiferromagnetic} critical point.  Figures \ref{gppsca} and \ref{gpplg}
are a similar plots for the Ising model on the simple cubic lattice
(for which $\rho_c = (3-e_c)/2 \simeq 1.997$),
and the NNE lattice gas, respectively.

\begin{figure}[!htb]
\caption{(Color online) Ising model, square lattice: $s''(\rho)$ versus
$\rho$ for system sizes $L=20$, 40, 80, 120, and 160.  The minimum in $|s''|$ grows sharper
with increasing system size.
}
\label{spp2d}
\end{figure}

\begin{figure}[!htb]
\caption{(Color online) Ising model, simple cubic lattice: $s''(\rho)$ versus
$\rho = n/L^3$ for system sizes $L=16$, 24, and 32.  The three curves are indistinguishable
at this scale.
}
\label{gppsca}
\end{figure}

\begin{figure}[!htb]
\caption{(Color online) NNE lattice gas, square lattice: $s''(\rho)$ versus
$\rho = N/L^2$ for system sizes $L=20$, 40, 80, and 120.
}
\label{gpplg}
\end{figure}

\subsection{A note on restricted sampling}

Our method, as noted, samples the entire range of energies, and while this
is advantageous in some circumstances, one might inquire whether the sampling
might be restricted or at least concentrated in a region of particular
interest, for example, to those energies that occur with a significant probability in the
critical region.  In Wang-Landau sampling, imposing fixed limits on the
sampling range has been found to distort the estimates for
$\Omega(E)$ in certain cases \cite{adaptive-window,nnelg}.  We therefore consider a {\it smooth}
preference in sampling, in which the acceptance probability, Eq. (\ref{paccept})
is modified so:

\begin{equation}
p_a({\cal C'})= \min\left[ \frac{\Omega(E) f(E)}{\Omega(E') f(E')},1 \right].
\label{pacceptf}
\end{equation}

\noindent Now the probability of visiting energy level $E$ is $\propto 1/f(E)$,
and the formula for updating $\Omega(E)$ becomes,

\begin{equation}
\Omega_{j+1} (E) = \frac{H_j (E)}{\overline{H}_j} f(E) \Omega_j (E).
\label{refinef}
\end{equation}

\noindent In principle, one can penalize visits to ``uninteresting" regions using larger values
of $f$, which are then compensated for in the modified updating formula.  Since a
greater fraction of the sampling occurs in the region of interest, one expects
to obtain better statistics there, for the same total number of lattice updates.

We set $f=1$ in the region of interest, that is, for $n \geq n_{min}$.  In one series
of studies, $f = \exp[\alpha (n_{min} - n)]$ for $n < n_{min}$, while in the second we
set $f = \exp[\alpha (n_{min} - n)^2]$.  Here $\alpha$ is a parameter that controls
the sharpness of the cutoff in sampling.  In the linear case, using $n_{min} \approx L^2$ (i.e.,
half the maximum bond number), and $\alpha$ in the
range 0.1 - 0.5, we obtain distributions $\Omega(n)$ (for $L=40$ and 80),
that differ significantly from those obtained
with uniform sampling, even on the region with $f=1$.  In the second set of studies we chose $\alpha$
such that $f(n=0) = e^{-5} \simeq 0.007$; $f$ grows quite slowly in the vicinity of $n_{min}$.
In this case the estimated number of configurations does not
exhibit obvious distortions, but we find that the susceptibility and specific heat, evaluated near
$T_c$, are significantly greater than the values obtained with uniform sampling.
The relative differences grow with the system size.
For example, the maximum susceptibility $\chi_{max}$ is found to be 0.6\% (1\%) higher
than that obtained using unrestricted sampling, for $L=40$ (80).
This occurs despite the fact that at $T_c$,
the the probability $e^{-\beta E} \Omega(E)$ associated with bond numbers
$n \leq n_{min}$ is $\leq 10^{-30}$, so that the contribution to thermal averages
due to this range of $n$ values is completely negligible.
We conclude that restricting sampling in this manner does not improve the quality of
the results, and in fact appears to cause systematic errors in thermal averages.

\section{Summary}

We have devised an entropic sampling method that covers the entire range of energies, and
yields results of good precision for a modest expenditure of computer time.
For the two-dimensional Ising model, we obtain the critical temperature to within
about 0.01\%, the exponent ratios $\gamma/\nu$ and $\beta/\nu$
to within 1\%, and excellent agreement with
theoretical results for the position and value of the specific heat and susceptibility maxima.
Very good results for the Ising model on the simple cubic lattice, and for the
lattice gas with nearest-neighbor exclusion, are also obtained.
The method is relatively simple,
and avoids sampling in ``windows" with the attendant problems of patching together
the results, and of possible distortions at the boundaries.

The sizes used in the present study are relatively modest, and indeed very
large sizes are not required to obtain precise results for the Ising model.  (One benchmark
study of the three-dimensional Ising model \cite{blote95}, for example, uses $L \leq 40$.
As the latter work illustrates, precise results for a large set of modest system sizes, coupled
with finite-size scaling analysis, may be just as useful as simulations of much larger
systems.)  In the present study we observe a reduction in quality (as reflected in the
relative uncertainty of thermodynamic variables), with increasing system size.  We believe
that an enriched set of initial configurations, permitting more thorough sampling of
configuration space,
will help to maintain precision as one increases the system size.  This is an important
issue for future work.

Key features of our {\it tomographic} sampling method are: pooling results starting from
very different initial configurations before updating the estimate for $\breve\Omega$,
and using the results for a given system size to generate an initial estimate
for the next larger size.
The method furnishes globally accurate results, as well as
precise estimates for critical properties, using a single set of parameter-free
simulations, thereby opening new possibilities for the application of entropic sampling.

\noindent \section*{Acknowledgments}

We are grateful to CNPq, and Fapemig (Brazil) for financial support.


\begin{thebibliography}{999}
\bibitem{lud} By a `lattice update' we mean
one attempted spin flip per site, or $L^d$ attempts, using a single-spin-flip
algorithm.
\bibitem{note_luds} We use
$N_U=2 n \times10^6$ at the $n$-th iteration.
\end{thebibliography}
\end{document}